\documentclass[nofootinbib,reprint,amsmath,amssymb,aps]{revtex4-2}
\usepackage{amsmath,pgf,float,appendix, hyperref,scalerel,amssymb}
\usepackage{footnote,stackengine,tikz,
    graphicx,subcaption,bm,dcolumn,tabularx}
\newcommand{\beq}{\begin{equation}\begin{aligned}}
\newcommand{\eeq}{\end{aligned}\end{equation}}
\def\msquare{\Omega}

\begin{document}
\title{On the Colloidal Phase of the Homogeneous Electron Fluid}
\author{Tom Banks and Bingnan Zhang}%
 \email{bingnan.zhang@rutgers.edu}
\affiliation{Department of Physics and NHETC
Rutgers University, Piscataway, NJ 08854}

\begin{abstract} We provide semi-rigorous arguments that the Homogeneous Electron Fluid (HEF) has a colloidal phase separating the Wigner Crystal from the high density fluid phase. Near the crossover between crystal and fluid ground state energies, the argument is quite general and valid for practically any\footnote{The argument depends on an inequality between the number of semi-classical colloidal states between which the Hamiltonian has $o(1)$ matrix elements, and $L^{2d -2}$ , where $L$ is the linear size of the system.  This inequality is extremely plausible, but we do not have a proof of it.} quantum transition between a crystal and a more amorphous phase.  In this regime, the colloid is a {\it gel} and its "Goldstone" modes are flows of irregular fluid droplets separated by crystalline walls.  A metal insulator transition occurs when a single bubble of fluid spans the entire system. Beyond this transition the colloid is a {\it sol} and its properties depend on the existence of stable finite crystallites with negative surface tension.  If these exist, the sol phase has lower energy than the homogeneous fluid.   Finally, we suggest a Landau mean field theory for the second order quantum phase transition between the fluid and sol phases.

\end{abstract}
\maketitle

\section{Introduction}
\par  
In previous work\cite{heginscol} on the Homogeneous Electron Fluid (HEF) we provided evidence for a colloidal phase intervening between the Wigner Crystal and high density homogeneous phases of the model in two and three dimensions.  The colloidal phase of the HEF was first suggested by Kivelson and Spivak\cite{kivspiv} in two dimensions.  They argued that colloidal ground states (which they called microemulsions) were lower in energy than the homogeneous fluid.  They also conjectured a second order phase transition between the colloidal and homogeneous phases.  Kivelson and Spivak\cite{kivspiv} argued that in two dimensions such negative surface tension objects always exist, at least in the form of stripes.  We argue that the existence of stripes also implies the existence of finite negative tension elliptical crystallites and the detailed competition between the finite crystallite phase and striped phases is difficult to calculate. Kivelson and Spivak also considered similar amorphous phases. We also provide weaker arguments that finite stable crystallites exist in three dimensions, arguing that the well known negative compressibility of the HEF gives a negative contribution to the surface tension.  

Subsequent to the work of\cite{kivspiv}, studies based on numerical/analytical resummations of high density perturbation theory in three dimensions\cite{gapless}\cite{haule} indicated the existence of a gapless state with non-zero wave number, a signal of a second order quantum phase transition to a non-homogeneous state\footnote{We thank Prof. Haule for conversations about his work, in which he indicated his belief that the results implied an instability to formation of an inhomogeneous phase, signalled by the appearance of a gapless excitation with non-zero wave number.  The text of the paper does not emphasize this interpretation of its results.}, at densities far above that at which the Wigner Crystal forms.  
In\cite{gapless}, Takada proposed an approximate set of Schwinger-Dyson equations, which led to a gapless excitation in the charge density channel (where Fermi liquid theory predicts only gapped plasmons), at a density much higher than the 
crossover between crystalline and homogeneous fluid phases found in Quantum Monte Carlo simulations.  Below this density, one finds both negative compressibility and a negative static dielectric constant.  The latter is incompatible with a homogeneous ground state.   

We note that the subject of colloidal phases of condensed matter systems does not begin with\cite{kivspiv}, and there is an extensive literature on the subject in particular materials\cite{collliterature}.  Kivelson and Spivak were the first to argue for the existence of phases with no translation symmetry in the jellium model, and experimental systems that are well modeled by it.

Our first indication of the existence of the colloidal phase came from a study of possible bubble nucleation instantons that destabilize a meta-stable crystalline state\cite{heginscol}.  We argued that elastic forces in the crystal would prevent the completion of the phase transition and instead produce a ground state that was a superposition of quantum states containing bubbles of fluid embedded in the crystal, with their growth and motion impeded by the strain energies involved.  In a region where the probability of bubble nucleation was very small, the stabilized configurations would have macroscopic bubbles, so that the superposition was just a classical probability distribution.   We had hoped to turn this intuition into a controlled calculation in the large $N$ expansion, where the Coulomb field coupled to the electron density operator becomes a semi-classical variable, but we showed that both the Wigner crystal and colloidal phases of the model occur at densities of order $1/N$, where straightforward large $N$ calculations break down.

 We will present a moderately rigorous variational argument in the regime just below $r_{WC}$ that the stable "homogeneous" ground state in that regime is in fact a coherent superposition of states with a non-homogeneous density profile, with macroscopic features. These are the colloidal states.
As usual in the quantum mechanics of macroscopic objects, actual measurements will always find some particular colloidal configuration, with the quantum wave function just telling us the probability for finding each configuration.  We further argue that while the individual colloidal states are not eigenstates of the Hamiltonian, their time evolution is dominated, for practically achievable measurements, by classical evolution of the constituents of the colloid.  In this low density {\it gel} regime, this means slipping of crystalline boundaries over fluid bubbles, and possible combination of bubbles.
The argument we will present is in fact applicable to the ground state of any quantum system that has a regime where a truly homogeneous fluid state has lower energy density than a crystalline state.   That is, we expect that most systems with these properties have a range of control parameters where the true ground state is a colloidal superposition.  We emphasize that our analysis applies only to quantum liquid/crystal/colloid transitions, in the neighborhood of the control parameter where the homogeneous fluid state first has lower energy than the homogeneous crystal.

If one considers such a low density colloidal phase for the HEF and imagines its evolution as the density is increased and the fluid phase becomes more and more favored energetically, one is led to the conclusion that there will be a metal insulator transition, which does not lead to a singularity in the ground state energy, when an infinite bubble of fluid is the dominant configuration in the colloid\footnote{We thank S. Kivelson for explaining this to us.}.  At higher densities, the colloid phase will be in a {\it sol} regime.  Typical configurations will be crystallites immersed in fluid.  One may ask why these crystallites don't melt.  The answer depends on the surface tension of the crystallites immersed in the fluid.

The sol regime exists if the crystallites have negative surface tension.  Indeed, a system with a meta-stable and unstable phase, each of which is invariant under an infinite subgroup of translations, will always have excitations where a region of radius $r$ of the higher energy phase forms.   If the surface tension of such a region is negative, then there will be a stable negative energy (relative to the fluid ground state), excitation of some size $r_0$.  This means that the homogeneous fluid is not the ground state, but rather that the ground state is a superposition of distributions of these excitations.   If $r_0$ is much larger than microscopic scales, the excitations themselves will be systems with many degrees of freedom.  In particular, if the meta-stable phase is crystalline, the crystallites will have many rotational levels with large values of angular momentum, and will be characterized by a non-zero wave vector $k_0$, and a particular space group.
They are thus a sort of negative energy soliton in the fluid phase of the system.  The true ground state is a superposition of semi-classical states with finite densities of crystallites immersed in the fluid.  The large crystallites will not behave like elementary particles with Bose or Fermi statistics.  Imagine such a crystallite in the fluid phase, scattering off a quasi-particle of the fluid.  Some of the energy transferred to the crystallite will go into exciting its many low energy levels rather than moving it.  The crystallites are thus distinguishable, and will behave like particles with Boltzmann statistics, to the extent that they behave like particles at all.

If we think about a mesoscopic/macroscopic crystallite, it can clearly have both odd and even numbers of fermions.  When the number is large fermi statistics is irrelevant and the excitations obey Boltzmann statistics.  As the surface tension goes to zero, the size of the crystallites shrink.  At some point we have to treat them as quantum particles.  As long as some of these particles are bosons, we will have Bose condensation if the gap goes to zero.  The crude estimate of the energy of these objects given by balancing surface and volume energies goes to zero, but this is certainly not an adequate argument.  
Our conjecture that the gap goes to zero is based entirely on the calculations of \cite{gapless}\cite{haule}.  We propose that the second order phase transition conjectured by\cite{kivspiv} and discovered in\cite{gapless}\cite{haule} could be interpreted as Bose condensation of crystallite solitons.  The essential characterization of the sol regime of the colloidal phase is the existence of negative surface tension excitations.

In\cite{heginscol} we attempted to argue that negative surface tension could be understood solely on the basis of repulsive Coulomb interactions at the surface of the crystal.  In the work to be reported in this paper we show that the situation is more complicated.  We'll argue that classical Coulomb forces alone do not give rise to negative surface tension, but that another source of negative tension is the negative compressibility of the homogeneous phase of the HEF, which has been found by a variety of other calculational methods.  Quantum Monte Carlo calculations with finite numbers of electrons have reported negative compressibility for many years\cite{QMC}. Some monographs\cite{italians} on the electron fluid tend to dismiss it as "an artifact of an unrealistic model", but negative compressibility has been measured in two dimensional HEFs\cite{eisenstein}.  The calculations of \cite{gapless}\cite{haule} find that it sets in precisely at densities below the point where the gapless excitation appears.  Furthermore\cite{gapless} finds that it is accompanied by a negative static dielectric constant, which is impossible in a stable homogeneous phase\cite{negdiel}.   
These results suggest that negative compressibility of the homogeneous fluid is in fact the sign of the existence of a more stable inhomogeneous phase.  We note however that the authors of\cite{lorenzana} use a mean field argument to argue that there is no instability.  The mean field theory of\cite{kivspiv} is more
sophisticated, and it is ambivalent, as we remain, about the existence of negative tension crystallites in three dimensions.   Our main reason for believing that the sol regime exists in three dimensions is the demonstration in\cite{gapless} that the dielectric constant is negative.   

As noted, in three dimensions we are unable to prove that the crystallite surface tension is negative, but only to exhibit both positive and negative contributions with similar sizes.  In two dimensions, the arguments of\cite{kivspiv} show that the negative "surface tension" is parametrically larger than the positive surface tension contributions, growing more rapidly than $R$, the radius of the crystallite, if the crystallite is sufficiently a-spherical.  In two dimensions a striped phase is always lower energy than the homogeneous fluid, but it also competes with sols made of a-spherical crystallites.   In three dimensions, Overhauser\cite{overhauser} has demonstrated that, in the Hartree-Fock approximation, a state with spin density wave order lies below the homogeneous state at sufficiently low densities, far above the WC transition.  There may also be charge density wave instabilities of the homogeneous fluid. It is therefore plausible that there are stable a-spherical crystallites in three dimensions as well, in which finite lengths of the density wave give rise to finite energy solitons with negative tension, floating in a homogeneous background.   

This paper consists of several independent sections, each of which addresses the existence of a colloidal phase in different ways.  
In Section \ref{gel} we present a variational argument that the phase at $r_S$ just below $r_{WC}$ is a colloid.  The argument uses only the existence of neutrally stable bubbles of fluid inside the meta-stable WC, translation invariance, and the fact that near $r_S = r_{WC}$ the bubbles are large and contain many electrons.   This argument is very general and applies to any would be first order quantum transition between crystal and fluid phases.  We exhibit an exponentially large (in the volume) set of quantum states 
whose expectation value of the Hamiltonian is larger than that of the homogeneous fluid in a box of volume $L^d$, by an amount of order $L^{d-1}$.  If the subspace of these states with Hamiltonian matrix elements $\langle m | H | n \rangle \sim o(1)$ has dimension $M$, with $M > L^{2d - 2} $, then random matrix theory tells us that there are superpositions of these states with lower energy than the fluid.  Since the space of states over which we can search for order $1$ matrix elements has dimension $e^{c L^d}$ it would seem that colloids should be the general rule for all {\it quantum} transitions between crystalline and more homogeneous states.  

In section \ref{meta} we prove a theorem that meta-stable states with the symmetries of the WC always exist if we put the system on a microscopically small lattice with periodic boundary conditions.  The lowest lying small excitations of these states are simply the motion of the crystal around the lattice (since there are many lattice sites per crystal unit cell).  This theorem does not prove the existence of a true crystal, with a peaked density distribution.  We use the variational approximation to estimate the value of $r_S$ at which a peaked distribution appears and find that it is reasonably small, so that the calculation should be reliable.  This symmetry argument does not prove absolute stability of periodic ground states, since a discrete translation invariant superposition of holes in the periodic configuration would have the same symmetry properties. However, the instability of these states is due to finite size holes,
so small enough regions of these crystalline configurations will be stable if their surface tension is negative.   An essential part of our argument is that the range of $r_S$ where the stability of a small crystallite is entirely determined by its surface tension, overlaps with the range in which the bulk compressibility of the fluid phase is negative.  The variational arguments in section\ref{meta}, combined with QMC calculations, and the estimates in\cite{haule}\cite{gapless}, provide evidence for this overlap.

In Section \ref{sol} we study the {\it sol} regime of the colloidal phase and attempt to argue that crystallites have negative surface tension.  The argument is incomplete in three dimensions.  Using the fact that the non-colloidal homogeneous phase has negative compressibility, we can find a negative contribution to the surface tension.  We cannot however find a definitive argument that the negative contributions dominate over other positive contributions that appear in our estimate.  In two dimensions, the arguments of\cite{kivspiv} show that the negative contributions dominate, parametrically in the radius of the crystallite as long as the crystallite has volume to surface ratio that grows slowly enough with size.

In section \ref{mean} we present a mean field theory describing the regime of densities just below the phase transition between the fluid and the colloid and show that it admits the sort of negative surface tension crystallites discussed above, in certain parameter regimes in which the compressibility of the fluid is negative.  The mean field theory is based on the Local Density Approximation (LDA) to the density functional.   It implies that the crystallites are bosonic quasi-particles near the transition, since they are excitations of the quantized density operator.  This is the same conclusion one reaches from the analyses of\cite{gapless}\cite{haule}, which find a gapless excitation in the charge density channel.

Section \ref{sum} contains a summary and conclusions.

\section{The Gel Regime}\label{gel}
\par
 We were originally led\cite{heginscol} to consider the colloidal phase by thinking about instanton (bubble nucleation) transitions from a meta-stable WC to the fluid state.  We argued that these transitions would lead to a {\it gel}, in which bubbles of fluid are trapped in the interstices of a crystal, instead of a homogeneous fluid.  The instanton description of this transition would become systematic in the $1/N$ expansion but in fact, the leading large $N$ density functional has no meta-stable crystal solution, so we were unable to provide rigorous evidence, even at large $N$, for the existence of the colloidal phase.  

Here we will attempt to make a direct variational description of the gel regime.  We will begin by assuming that the exact density functional in the density regime $n_0 > n_{WC}$, just above the transition to the stable WC, has a neutrally stable stationary point with the following form
\begin{equation} n_{bubble} (\vec{x}) = n_0 f(\frac{|\vec{x}-\vec{x_0}|}{R}) + \left(1 - f(\frac{|\vec{x}-\vec{x_0}|}{R})\right) n_C (\vec{x}) . \end{equation} 
This describes a spherical bubble of fluid with radius $R$ embedded in the meta-stable crystal.
$f(|\vec{x}|)$ is a smooth approximation to a step function that is $1$ on the interval $[0,1]$.  Here $\vec{x}$ is the vector position in space and  $|\vec{x}-\vec{x_0}|$ is the Euclidean distance from some fixed spatial point. We will have a configuration of this type for any choice of that point. $n_C (\vec{x})$  is the density expectation value in the meta-stable crystal state.   The mean field arguments of Kivelson and Spivak suggest that the bubble will have a net non-zero surface charge density.  This will be screened, by polarization of the crystal surrounding it.  The long range interactions between this bubble and another one far from it, will be mediated by phonons.

 The parameter $R$ will depend on $r_S$.  All eigenvalues of the functional Hessian of the density functional are positive at this bubble configuration except for $d + 1$ zero modes.  One zero mode corresponds to variation of $R$ and the derivative with respect to $R$ changes sign at $R(r_S)$.  The other zero modes are exact collective coordinates, which translate the center of the bubble. 
If $r_S$ is close to the transition to the stable WC phase, the negative energy density difference between the crystal and homogeneous fluid phases is small.  $R(r_S)$ is determined by balancing this against the positive surface tension of the fluid bubble
\begin{equation} V_dR^d\Delta\epsilon=A_{d-1}R^{d-1}\sigma_S\Rightarrow R\propto \frac{\sigma_S}{\Delta\epsilon}  \end{equation}  where $\sigma_S$ is the surface tension, $\Delta\epsilon$ is the energy density difference between the fluid and the crystal, $V_d$ is the volume of the unit d-ball, $A_{d-1}$ is the area of the unit (d-1)-sphere. Since $\Delta\epsilon$ is small, $R(r_S)$ is large in this regime and the bubbles contain of order $n_0 R^{d} (r_S) $ electrons.  The number of quantum states corresponding to a bubble of this size is $e^{b n_0 R^d }$, with $b \sim 1$, and many of these states have energy very close to that of the bubble, so that we are justified in thinking of the bubble as a classical object. 

As a consequence of the collective coordinates there are approximate multi-bubble stationary points in which we have a finite density of bubbles, centered at positions $ \vec{x_i}$ separated by distances $| \vec{x_i} - \vec{x_j}| \gg R(r_S)$.  These finite bubble density configurations have negative energy density relative to the infinite crystal.  We can lower the energy density of these configurations in many different ways, by choosing to expand the radii of some finite fraction of the bubbles, and perhaps contract others.  Eventually we get into the regime in which some of the inter-center distances are of order the bubble radii.  

Even before this happens, there is a qualitative difference between the energetics here and those of bubbles of one fluid phase inside another.   The low energy excitations of the crystalline phase are phonons.  Expanding a single bubble in an otherwise perfect crystal distorts the lattice, which is to say that the change in radius is accompanied by phonon excitations.  This can be seen by applying elasticity theory\cite{elasticity}:  The bubble can lower its energy as it expands because $dE/dR$ is negative for $R > R_C$ the critical radius.  However this imposes a strain on the crystal, and stress is proportional to strain.  The elastic energy, the product of stress and strain, is proportional to $(dE/dr)^2$ and would stablize the bubble if the strain were localized at the bubble wall.  By instead distributing the strain over a long wave length distortion of the crystal, we can reduce the energy and allow the bubble to expand.  The use of continuum elasticity theory is justified in this context because the bubbles and the crystalline walls separating them are large macroscopic objects near the crossover density between fluid and WC states.
\begin{figure}[H]
\centering
\includegraphics[scale=0.6]{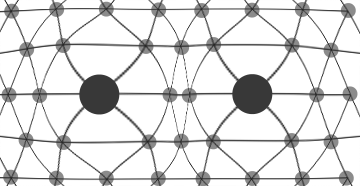}
\caption{The waves surrounding one bubble interact with those from another}
\label{bubble}
\end{figure}
 Thus, for a single bubble, we can make the phonon energy low by employing a radial wave of very long wavelength.  However, as bubbles approach each other the waves surrounding one bubble interact with those from another as shown in figure \ref{bubble}    .  With an infinite distribution of bubbles a locally stable state will not allow all of the bubbles to meet and collide. Rather, we will have low energy density configurations of $n(\vec{x})$ where large bubbles are separated by macroscopic walls of crystal.  

In particular, if we examine configurations of $n(\vec{x})$ on a torus of length $L$, for which the volume of space filled by fluid bubbles scales like $L^d$ while the crystalline volume is $L^{d - 1}$ each of these configurations would have the same energy density in the large $L$ limit as a configuration with $n(\vec{x}) = n_0$.  Any excess energy would scale like $L^{d - 1}$.

It is clear that the number of such meta-stable multi-bubble configurations of $n(\vec{x})$ with the same energy density is of order $e^{c L^d / V_{bubble}}$, where $V_{bubble}$ is the volume of a typical member of the bubble ensemble.  The bubbles will be of many sizes and not all of them will be spherical.  Non-spherical shapes come from bubbles that have collided and coalesced\footnote{Although we use dynamical language, there is no dynamics going on here.  We are simply describing the functional landscape of configurations of $n(\vec{x})$ that have the same energy density.} in the process of lowering the energy.   It is important to realize that each of these configurations corresponds to an exponentially large number of quantum states if we are in a regime where the bubbles contain $10^k$ electrons with $k > 2$ (say).   For large bubbles, like those we expect near the transition at $r_{WC}$, many of these states will be very close to the lowest energy state with the given density profile.  There will also be different density profiles that produce states of equal energy through order $L^{d-1}$ in the large $L$ expansion.

The matrix elements of the Hamiltonian between two close by configurations $n(\vec{x}), n^{\prime} (\vec{x})$ will be of order\footnote{The Gaussian formula for the overlap between two states with different density configurations is valid because the overlap is maximized when the density configuration is the same so we can make a power series expansion dominated by the quadratic term of the log of the overlap.  The specific form of the quadratic term is because this is proportional to the volume over which the densities are different.  Terms involving derivatives of the density will be subleading in the volume.} $e^{ - g \int d^d x (n(\vec{x}) - n^{\prime} (\vec{x}))^2}$ and even smaller for configurations that are more distant.  Close by configurations will include those where $n(\vec{x}) = n^{\prime} (\vec{x} + \vec{a}) $ with $a$ small compared to bubble sizes, and those where small distortions of a finite number of bubbles occur.  If we study the time dependence of a state with the minimum expectation value of energy for some particular $n(\vec{x})$ it will, to an excellent approximation, be described by classical time evolution of the density distribution $n(\vec{x},t)$, including flow and smooth distortions of the bubble shapes.  The true lowest energy state will be some superposition of all of these colloidal states, and will be invariant under continuous spatial translation.  

However, as in the description of any macroscopic system in quantum mechanics, actual coarse grained measurements at a fixed time will only be sensitive to a particular value of $n(\vec{x})$, and of other macro-observables like the spin density,  so despite the mathematical translation invariance of the true ground state, the system will appear amorphous to any reasonable probe.  This is spontaneous breaking of translation invariance, leaving not even a discrete subgroup unbroken.  The slow movement and shape changes of bubbles are the Goldstone modes for this symmetry breaking pattern.  Infinitesimal Goldstone modes can be quantized.   Since the individual gel configurations are completely inhomogeneous, the Goldstone modes are not quasi-particles with a fixed dispersion relation or uniform properties throughout the system.  In the crystalline portion of the gel they will be phonons, but scattering off the fluid interfaces will give them short lifetimes.  We believe that the problem of low frequency transport in this phase is essentially classical.  One must solve the quantum mechanics better than we have, in order to determine the most probable configurations $n_i (\vec{x})$ at each density $n_0$, and to determine the classical field equation that controls how they vary with time.  

If we consider matrix elements of the Hamiltonian between this huge set of quantum states they will form diagonal bands, with matrix elements of order $1$ between states with very similar density expectation values and very small matrix elements between those with large relative distortions of $n(\vec{x})$.   Call the width of the bands with order one matrix elements, $D$.   A random $D \times D$ matrix has eigenvalues distributed by the Wigner semi-circle law between $\pm \sqrt{D}$.   Then if $\sqrt{D} > L^{d - 1}$  the lowering of energy due to superposition of colloidal states will overwhelm the difference in the the value of the density functional between the truly homogeneous state and any individual member of the colloidal superposition.  The state with minimum energy for constant density expectation value will actually be a colloid.

While we do not have a rigorous argument that $D > L^{2(d - 1)}$, the space of states with energy expectation value exceeding that of the homogeneous state by $\sim L^{d - 1}$ is exponentially large in the volume.  It seems plausible that the subspace where the Hamiltonian is well approximated by a Wigner random matrix with order one matrix elements is also huge, although not exponential in the volume.

This analysis shows that, although the density expectation value of the "true quantum ground state" of the HEF in the regime just above the transition to the WC is uniform, the actual phenomenology of this state will be that of a gel.   Any measurement will find a non-uniform distribution of $n(\vec{x})$, which will change with time on a scale long compared to the natural microscopic time scales of the system and be governed by a classical equation of motion with tiny statistical fluctuations.   

We believe that this is a physically rigorous argument (though a better lower bound on $D$ would certainly be needed for true rigor) that the homogeneous phase whose energy crosses below that of the WC at the transition points estimated by Quantum Monte Carlo (QMC)\cite{QMC}, calculations is actually colloidal.  Note that QMC calculations cannot, even with reasonable extrapolations from the current state of the art, distinguish between the colloid and a truly homogeneous phase.  The number of electrons in a QMC calculation is much smaller than the number in a single bubble of the gel in the regime near $r_{WC}$.  

A possible strategy for numerical investigation of the gel regime would be to use QMC calculations to determine wave functions for small finite volumes of fluid and crystal, and then combine these to cover the bubbles and crystalline walls in a fixed density profile of the gel.  As complete amateurs at such numerical simulations, this appears to the authors to be a task of insurmountable difficulty, but perhaps experts can find a way.

As $r_S$ is lowered, the neutrally stable bubbles from which we began our exploration of the density functional landscape get smaller and the probability that bubble mergers will be impeded by stressed crystalline walls decreases.  At some value of $r_S$ we will encounter a metal insulator transition where the probability of a fluid region stretching across the whole volume becomes of order $1$.  This does not alter the bulk thermodynamics of the system, but does affect the transport of charge and heat.  At some point, a better qualitative description of the system is a {\it sol} : crystallites immersed in fluid.   

The fate of crystallites is our next object of study.  We emphasize that the discussion of this section appears quite universal to first order quantum phase transitions between a crystal and a more fluid state of matter.  Such transitions are likely to always lead to gel states, for some range of control parameter.   The existence of a sol phase is not however universal.  The crystallites in the sol can shrink rapidly, melting into a homogeneous fluid.  We will argue in the next section that the crucial criterion for the existence of a sol phase is negative surface tension, as first posited by\cite{kivspiv}.  We will also identify both positive and negative contributions to the surface tension, but we are not able to give a definitive argument that the negative terms dominate.  Kivelson and Spivak\cite{kivspiv} presented a mean field argument that this is indeed the case in two dimensions: the negative surface tension coming from transferring charge between crystallite and the surrounding fluid is proportional to $R{\rm ln}\ (R/R_0)$, and overwhelms positive contributions to the tension for large enough $R$. 

Crystallites only exist if sufficiently small regions of crystal are locally stable.  We will begin by establishing this fact by a symmetry analysis.  The question of global stability then comes down to the balance of positive volume energy (relative to the fluid) versus surface tension.  Negative tension ensures stability, and implies that a sol of crystallites has lower energy than the homogeneous state.

\section{Meta-stable States with Crystalline Symmetry}\label{meta}
\par
We argue that, in a finite periodic lattice version of the HEF, there are always metastable states with the symmetries of a Wigner Crystal.  This does not imply a true crystalline structure, in the sense that the ground state expectation value of the density operator has sharp peaks.  We will show that, in a simple variational approximation, sharp peaks in the density expectation value appear only for sufficiently low density.  Below that density, there are always metastable states that are truly Wigner Crystals.  At high density, the WC does not become unstable by developing a phonon mode with imaginary frequency.  It just becomes a periodic jelly, with no crystalline structure.  

Our Hamiltonian is a lattice approximation to the Homogeneous electron gas

\beq H = &\sum_{<i,j>} \psi^{\dagger} (i) t(i,j) \psi (j) \\&+ \sum_{i,j}(n (i) - n_0)G(i-j)(n(j) - n_0). \eeq $n(i) = \psi^{\dagger} (i) \psi (i) $ and we work in a sector in which the conserved charge $Q \equiv \sum_i n(i) = n_0 L^d $, where $d = 2,3$.  $- G$ is the Green's function of the lattice Laplacian on a toroidal lattice of side $L$, whose matrix is $t(i,j)$. $t(i,j)$ is nonzero only for nearest neighbor $i,j$. We work in length units where the lattice spacing is one, and for definiteness we can think of it as $10^{-3}$ Bohr radii.  Here and throughout this paper, we work with spinless electrons.  In the lowest order in small electron velocities, electron spin is an internal quantum number and this can be exploited to construct large $N$ approximations.  We expect many aspects of the physics to be independent of $N$, and we see no real difference in our arguments as $N$ is varied, apart from the fact that the transitions that are the main focus of this paper occur at densities of order $1/N$\cite{heginscol} and are inaccessible to straightforward large $N$ calculations.  In the interest of simplifying the equations we simply work with $N = 1$.  This means of course that we cannot study phases of the system with magnetic order.  We will continue to call our spinless fermions "electrons".  

The Hamiltonian is invariant under the $d$ commuting single lattice site translations $T_a$ and so the Hamiltonian can be diagonalized independently in sectors where the $T_a$ are diagonal, with eigenvalues $e^{i\phi_a}$.  These phases are the Bloch momenta of this problem. Let us consider sectors where each of these eigenvalues is an $M$th root of unity with $L \gg M \gg 1$.  These are states invariant under $T_a^M$.  Let's further restrict attention to states that are not invariant under any smaller power of the $T_a$.  We'll call this a {\it crystalline sector}, with unit cell of size $M$.  We introduce these sectors because the ground state in each sector is absolutely stable.
We can further refine our classification of states by specifying the space-group that leaves a given state invariant, but that will not be necessary for present purposes since we'll restrict our attention to the classic Wigner crystal structures that give lowest energy at very low density.  

Let $E_0 (\phi_a )$ be the ground state energy in the sector with a particular $Z_M^d$ eigenvalue, and $| s_0 (\phi_a) \rangle $ the corresponding state.   The time evolution of 
\begin{equation} | \psi_C (x_a ) \rangle = \sum_{\phi_a = 0}^{\frac{2\pi (M - 1)}{M}} e^{ i x_a \phi_a} f (\phi_a ) | s_0 (\phi_a) \rangle , \end{equation} is
\begin{equation} | \psi_C (x_a ) \rangle = \sum_{\phi_a = 0}^{\frac{2\pi (M - 1)}{M}} e^{ i x_a  \phi_a - i E_0 (\phi_a)t} f (\phi_a ) | s_0 (\phi_a) \rangle . \end{equation}   We choose $f$ so that only relatively low Bloch momentum modes are supported, but at large $M$ this still allows the wave function to be localized around $x_a$ .  At large $t$ as long as $E_0$ is a smooth function of $\phi_a$ this will just lead to a translation of $x_a$ with a velocity determined by the stationary phase point $x_a = t \partial_a E_0 $.  Since $M$ is large, we expect the energy for small $\phi_a$ to behave like ${\phi_a C^{ab} \phi_b}{M^d}$ so the period $M$ structure will move slowly around the torus.
In the limit $L \gg M \rightarrow \infty$ it will approach a meta-stable periodic ground state, spontaneously breaking the microscopic lattice translation symmetry.  

We believe that this analysis shows that the structure is meta-stable, but it does not preclude the existence of a localized excitation of the periodic structure, with lower energy.  Indeed, if the lattice is large enough we can simply insert a critical bubble of homogeneous fluid into  the crystal. We can superpose such states, to make a state invariant under the crystal subgroup.  In the symmetry protected sectors with fixed transformation properties under $T_a$, we would have superpositions of states where the localized excitation is at different positions in the crystal.  In our construction of crystallites in the next section, we have to take the size and shape of the crystallite small enough that the critical bubble does not fit into it.  The detailed analysis of this constraint depends on the shape and boundary configuration of the crystallite.

To summarize, symmetry analysis on small toroidal lattices proves that there are small objects with crystal symmetry, which would be stable when immersed in an otherwise homogeneous fluid if they have negative surface tension.  We will now argue that, starting around $r_S = 4$, these states actually have sharp crystalline peaks in their density expectation value.

\subsection{Variational Analysis of Crystalline Structure}
\par
  We will use a variational approximation to make these estimates.  For those values of $r_S$ for which our calculation indicates a true crystalline structure, we can view our variational wave function as a particular Hartree-Fock wave function for which the exchange energies are very small because individual electron wave functions are localized at crystalline sites whose separation is much larger than the width of the single electron wave functions.  Small perturbations of these wave functions will change the detailed shape of the single electron wave functions but not have a dramatic impact on their width.  Thus, our variational approximation would lead to a local minimum of the Hartree-Fock functional with a true crystalline density profile.  

 The Hartree-Fock approximation is accurate at high density in the homogeneous phase, and we see no reason to doubt that it is similarly accurate in states with crystal symmetries and high density.  We find that the density profile of a state with crystal symmetry begins to show peaks at a relatively high density.   Combining this information with the known crystal structure of the ground state at very low density, and the meta-stability analysis of the previous sub-section, we have evidence for the existence of a true crystal state, that is meta-stable for all densities above some critical value $r_{C}$.  $r_{C}$ is not sharply defined because the definition of "sharply peaked density expectation value" is inevitably somewhat subjective.  In our variational approximation we will find $r_{C}  \sim 4$ in both 2D and 3D.

Although our argument for meta-stability used the symmetries and finiteness of the lattice model, we will perform the variational calculations in the continuum.  Since the values of $r_S$ at which we find true crystal structure correspond to $M \sim 10^4$ for a lattice spacing $10^{-3}$ of the Bohr radius, and even larger for finer lattices, the lattice and continuum variational calculations will be very close.  In addition, rather than trying to do an exact variational calculation, we will content ourselves with a Gaussian ansatz for the single particle wave functions around each lattice site.  This ansatz becomes exact in the {\it low density} limit.  We use it at relatively high densities because it establishes the preference for crystalline structure rather than a periodic gelatine.  Our intention is not to find quantitative estimates for the lowest energy crystalline state, but only to establish the qualitative preferences for sharp crystalline peaks.

\subsection{Gaussian variation in 2D}\par
We treat electrons as spinless fermions, and put Gaussian wave packets on the triangular Wigner crystal lattice. This approximation is good when the overlap between nearby wave packets is small. Define $\rho(\vec{x})=\sum_n \rho_n(\vec{x})$ where $\rho_n(\vec{x})$ is the density distribution of the $n_{th}$ electron, which is a Gaussian. $\rho(\vec{x})$ forms a triangular lattice, so $\rho(\vec{x})=\frac{1}{L^2}\sum_{\vec{k}\in D}\rho(\vec{k}) e^{-i\vec{k}\cdot \vec{x}}$, where $D$ is the dual lattice,$L^2$ is the size of the whole space.Define $\msquare$ as the volume of a unit cell in coordinate space. $N=\frac{L^2}{\msquare}$ is the electron number.
For $\vec{k}\in D$ we have 
\begin{equation}
\begin{aligned}
\rho(\vec{k})&=\sum_n \rho_n(\vec{k})=\sum_n \int d^2x \rho_n(\vec{x})e^{i\vec{k}\cdot \vec{x}}\\&=\sum_n \int d^2 y \rho_n(\vec{y}+\vec{R_n})e^{i\vec{k}\cdot (\vec{y}+\vec{R_n})}\\
&=\sum_n \int d^2y \rho_0(\vec{y})e^{i\vec{k}\cdot \vec{y}}=N\rho_0(\vec{k})
\end{aligned}
\end{equation}
Denote $G_k=\frac{2\pi}{k}$ as the Coulomb potential in momentum space, $n_0$ as the density of the background charge. 
\begin{equation}
\begin{aligned}
\frac{1}{2}\int d^2x\int d^2y \frac{\rho(\vec{x})\rho(\vec{y})}{|\vec{x}-\vec{y}|}&=\frac{1}{2L^2}\sum_{\vec{k}\in D}|\rho(\vec{k})|^2
G_k\\&=\frac{N^2}{2L^2}\sum_{\vec{k}\in D}|\rho_0(\vec{k})|^2G_k\\
&=N\frac{1}{2\msquare}\sum_{\vec{k}\in D}|\rho_0(\vec{k})|^2G_k
\end{aligned}
\end{equation}
\begin{equation}
\begin{aligned}
V_{ee}=&\frac{1}{2}\int d^2x\int d^2y \frac{\rho(\vec{x})\rho(\vec{y})}{|\vec{x}-\vec{y}|}\\&-\frac{1}{2}\int d^2x\int d^2y \sum_n\frac{\rho_n(\vec{x})\rho_n(\vec{y})}{|\vec{x}-\vec{y}|}\\
=&N\frac{1}{2\msquare}\sum_{\vec{k}\in D}|\rho_0(\vec{k})|^2G_k-\frac{N}{2L^2}\sum_{all\ \vec{k}}|\rho_0(\vec{k})|^2G_k\\
=&N\frac{1}{2\msquare}G_{k=0}+N\frac{1}{2\msquare}\sum_{\vec{k}\in D,\vec{k}\neq 0}|\rho_0(\vec{k})|^2G_k\\&-\frac{N}{2L^2}\sum_{all\ \vec{k}}|\rho_0(\vec{k})|^2G_k
\end{aligned}
\end{equation}
\begin{equation}
\begin{aligned}
V_{eb}=&-\int d^2 x\int d^2 y \int\frac{d^2 k}{(2\pi)^2}\rho(\vec{k})e^{-i\vec{k}\cdot \vec{x}}\\&\ \cdot\int \frac{d^2p}{(2\pi)^2}G_p e^{-i\vec{p}\cdot (\vec{x}-\vec{y})}n_0\\
=&-\int\frac{d^2k}{(2\pi)^2}\int\frac{d^2p}{(2\pi)^2}(2\pi)^2\delta^2(\vec{p})(2\pi)^2\delta^2(\vec{k}+\vec{p})\\&\ \cdot\rho(\vec{k})G_pn_0\\
=&-\rho_{\vec{k}=0}n_0G_{p=0}=-N\frac{N}{L^2}G_{p=0}=-\frac{N}{\msquare}G_{p=0}
\end{aligned}
\end{equation}
\beq
V_{bb}&=\frac{1}{2}n_0^2\int d^2x\int d^2y \int\frac{d^2k}{(2\pi)^2}G_ke^{-i\vec{k}\cdot(\vec{x}-\vec{y})}\\&=\frac{1}{2}n_0^2G_{k=0}L^2=N\frac{1}{2\msquare}G_{k=0}
\eeq
Sum up
\beq
V_{ee}+V_{eb}+V_{bb}=&\frac{N}{2\msquare}\sum_{\vec{k}\in D,\vec{k}\neq 0}|\rho_0(\vec{k})|^2G_k\\&-\frac{N}{2L^2}\sum_{all\ \vec{k}}|\rho_0(\vec{k})|^2G_k
\label{eq:1}
\eeq

As indicated above, we use the 2D Gaussian wave function $\phi_0(\vec{x})=\frac{1}{\sqrt{2\pi\sigma^2}}e^{-\vec{\vec{x}}^2/4\sigma^2}$. We can calculate the kinetic energy per electron
\begin{equation}
T=\int d^2x \phi_0(\vec{x})\frac{-\nabla^2}{2}\phi_0(\vec{x})=\frac{1}{4\sigma^2}
\end{equation}
and according to equation \ref{eq:1} the potential term is
\begin{equation}
V=\frac{1}{2\msquare}\sum_{\vec{k}\in D,\vec{k}\neq 0}e^{-\sigma^2\vec{k}^2}\frac{2\pi}{k}-\frac{1}{2}\int\frac{d^2k}{(2\pi)^2}e^{-\sigma^2\vec{k}^2}\frac{2\pi}{k}
\end{equation}
One cannot replace the discrete sum in the first term with an integral, because that only cancels the second term. We can do the summation numerically. Appendix \ref{C} is the plot of $\epsilon=T+V$ for several different $r_S$. Note that we should impose $\sigma\leq r_d/2=\sqrt{\frac{\pi}{4\sqrt{3}}}r_S$ where $r_d=\sqrt{\frac{\pi}{\sqrt{3}}}r_S$ is the distance between two lattice points on the two dimensional triangular lattice, or else there will be too much overlap between two adjacent electrons that the Hartree approximation is no longer valid. Indeed, although we are in principle applying the Hartree-Fock approximation, the region of interest consists of those values of $r_S$ for which the interference terms between electrons on different lattice sites are negligible: sharp crystalline peaks imply that exchange corrections are small. A saddle point appears below $r_d/2$ when $r_S\geq 4$. 

\subsection{Gaussian variation in 3D}
\par The analog of equation (\ref{eq:1}) in 3D is 
\begin{equation}
V_{total}=\frac{N}{2\msquare}\sum_{\vec{k}\in D,\vec{k}\neq 0}|\rho_0(\vec{k})|^2G_k-\frac{N}{2L^3}\sum_{all\ \vec{k}}|\rho_0(\vec{k})|^2G_k
\end{equation}
where $D$ is the dual lattice of the bcc Wigner crystal lattice. We use the Gaussian wave function  \begin{equation} \phi_0(\vec{x})=\frac{1}{(2\pi\sigma^2)^{3/4}}e^{-\vec{x}^2/4\sigma^2}, \rho_0(\vec{x})=\frac{1}{(2\pi\sigma^2)^{3/2}}e^{-\vec{x}^2/2\sigma^2}\end{equation} The energy per electron is 
\begin{equation}
\begin{aligned}
\epsilon &=T+V\\&=\frac{3}{8\sigma^2}+\frac{1}{2\msquare}\sum_{\vec{k}\in D,\vec{k}\neq 0}|\rho_0(\vec{k})|^2G_k-\frac{1}{2L^3}\sum_{all\ \vec{k}}|\rho_0(\vec{k})|^2G_k\\&=\frac{3}{8\sigma^2}+\frac{1}{2\msquare}\sum_{\vec{k}\in D,\vec{k}\neq 0}e^{-\sigma^2k^2}\frac{4\pi}{k^2}-\frac{1}{2}\int\frac{d^3k}{8\pi^3}e^{-\sigma^2k^2}\frac{4\pi}{k^2}
\end{aligned}
\end{equation}
\begin{equation}
\begin{aligned}
\frac{\partial\epsilon}{\partial\sigma^2}&=
-\frac{3}{8\sigma^4}-\frac{2\pi}{\msquare}\sum_{\vec{k}\in D,\vec{k}\neq 0}e^{-\sigma^2k^2}+2\pi\int \frac{d^3k}{8\pi^3}e^{-\sigma^2k^2}
\end{aligned}
\end{equation}
When $\sigma\ll r_S$ we have $\sigma k_D\ll 1$, where $k_D$ is the lattice spacing of the dual lattice $D$. We can replace the discrete summation with an integral
\begin{equation}
\begin{aligned}
\frac{\partial\epsilon}{\partial\sigma^2}&=
-\frac{3}{8\sigma^4}-\pi\int \frac{d^3k}{8\pi^3}e^{-\sigma^2k^2}+\frac{2\pi}{\msquare}+\pi\int \frac{d^3k}{8\pi^3}e^{-\sigma^2k^2}\\
&=\frac{2\pi}{\msquare}-\frac{3}{8\sigma^4}
\end{aligned}
\end{equation}
The saddle point is at $\sigma=(\frac{3\msquare}{16\pi})^{1/4}=r_S^{3/4}/\sqrt{2}$.\par
When the condition $\sigma\ll r_S$ is not satisfied, numerical summation is needed. Appendix \ref{D} is a plot of $\sigma/r_d-\epsilon$, where $r_d=\sqrt{3}(\frac{\pi}{3})^{1/3}r_S$ is the distance between two lattice points on the bcc Wigner crystal lattice. Again, a saddle point that satisfies $\sigma<r_d/2$ appears when $r_S\geq 4$.

\section{Negative Surface Tension and The Sol Regime}\label{sol}

Consider a connected region $D$ surrounding some point in space whose surface area scales like $R^{d-1}$ and whose volume scales like $R^p$ with $d - 1 < p \leq d$ for $d = 2,3$.  $D$ has an ellipsoidal shape. We also consider a shell of thickness $\Delta$ surrounding the region $D$. We will call the region outside the shell, $\bar{D}$.
In order to build a model for a crystallite occupying the region $D$, we will choose a variational wave function of the form
\beq \psi = \sum_{perm} (- 1)^P \psi_C (y_i) \tilde{\psi}_F (\tilde{y_j}) \psi_F (x_k) . \eeq
The functions $\psi_C$ and $\tilde{\psi}_F$ are chosen to vanish exponentially with a width $\delta$ whenever any of their arguments are outside of $D$ while $\psi_F$ has a similar exponential vanishing outside of $\bar{D}$.
The sum over permutations enforces Fermi statistics.

We work in the regime of $r_S$ where our previous arguments indicate the existence of a stable state whose density has sharp crystalline peaks, on a toroidal lattice of small enough size.  $\psi_C$ is the wave function of the stable crystal at that value of $r_S$ except right near the boundary of $D$, where it is modified by the exponential falloff.  Similarly $\psi_F$ will be the wave function of the homogeneous fluid modified by exponential falloff in the interior of the outer boundary of the shell surrounding $D$.  We will also insist that $r_S$ is in the regime where QMC calculations indicate that the fluid has negative compressibility.  QMC data show that the energy density of the fluid is {\it negative} in that regime\cite{italians}.

Our trial wave function excludes electrons from the shell region, which implies that the electron charge density inside or outside of the shell (or both) is larger than the background density $n_0 = \frac{c_d}{r_s^d} $.  We will choose the excess charge, whose magnitude is $n_0 V_{shell}$ to lie entirely inside $D$ and the wave function of the extra electrons is $\tilde{\psi}_F$.  

The expectation value of the HEF Hamiltonian in any state can be written as
\beq \langle H \rangle &= \langle K \rangle + \frac{1}{2} \int d^d xd^d y\ \frac{(\langle N(x)\rangle - n_0)(\langle N(y)\rangle - n_0)}{|\vec{ x} - \vec{y}|}\\& + \frac{1}{2} \int d^d xd^d y\ \frac{\langle N(x)N(y)\rangle_C}{|\vec{ x} - \vec{y}|}. \eeq
The kinetic energy $K$ and electron density operator $N(x)$ are one body operators and their expectation values receive separate and additive contributions from the regions $D,\bar{D}$ and vanish in the shell.  The connected correlator of the product of density operators is short ranged because of Debye screening, and also vanishes when one of the points is in the shell.  We will neglect its contribution when $\vec{y}$ is in $D$ and $\vec{x}$ in $\bar{D}$.  We will take $R \gg \Delta$, which for both $\vec{x}$ and $\vec{y}$ in D, implies that the contribution to this connected correlator from the electrons in $\tilde{\psi}$ is negligible compared to the crystalline contribution.

The expectation value of $N(\vec{x})$ vanishes when $\vec{x}$ is in the shell and equals $n_0$ when $\vec{x}$ is in $\bar{D}$.  For $\vec{x}$ in $D$ it can be decomposed into contributions from the crystalized electrons and the additional electrons described by $\tilde{\psi}_F$.  
\beq \langle N(\vec{x})\rangle = n_C (\vec{x}) + \tilde{n} (\vec{x}) . \eeq
Although $n_C \neq n_0$ pointwise, they do cancel when averaged over a large enough crystal. Denote the background charge in the three regions as $n_{0D},\ n_{0S},\ n_{0\bar{D}}$, so $n_0=n_{0D}+n_{0S}+n_{0\bar{D}}$. The Hartree terms involving $[n_C (\vec{x}) - n_{0D}][n_C (\vec{y}) - n_{0D}]$ contribute to the energy of the meta-stable crystallite.  
 
 With these approximations we can write
\beq \langle \psi | H | \psi \rangle &= \epsilon_C V_D + \epsilon_F V_{\bar{D}} \\&+  \frac{1}{2}\int d^d xd^d y\ \frac{(\tilde{n}(\vec{x})-n_{0S})(\tilde{n}(\vec{y})-n_{0S})}{|\vec{ x} - \vec{y}|} \\& +  \int d^d xd^d y\ \frac{(n_C(\vec{x})-n_{0D})(\tilde{n}(\vec{y})-n_{0S})}{|\vec{ x} - \vec{y}|} \\&+t A_D + \bar{t} A_{\bar{D}}, \eeq The next to last term is a positive surface tension coming from the part of the expectation value of the kinetic term concentrated near the boundary of the shell.  The coefficient $t$ is of order $\delta^{-1}$ in Bohr-Rydberg units.  We'll discuss the analogous contribution from the boundary of $\bar{D}$ below.

Apart from the volume energies of the crystal and fluid, and the two surface tension terms, the expectation value consists of self Coulomb repulsion between the charge densities in $D$ and in the shell and Coulomb attraction between the shell and the charge in $D$.  $n_C $ is not a variational parameter.  It is fixed by the wave function of the crystal.  $\tilde n(\vec{x})$ can be adjusted to minimize the energy by concentrating it near the boundary of $D$, maximizing the Coulomb attraction to the positive charge in the shell.  Furthermore, for large $R$, a boundary charge density will interact with fewer multipole moments of the crystalline charge density.  The same will be true of the interaction of the crystal with the positive charge in the shell.  Effectively we have a distribution of dipoles surrounding the crystallite.  This has a positive energy, which scales with a power of $R$ less than $d - 1$ for both $d = 2,3$ if the crystallite is sufficiently a-spherical.  It does not contribute to or compete with the surface tension.

Finally we turn to the coefficient $\bar{t}$ coming from the surface of $\bar{D}$.  Since we are in a regime of negative compressibility, this is negative.  The size of the surface area depends on $\Delta$.  In order to determine the optimal value of $\Delta$ we subtract off the energy of the homogeneous fluid phase, since we're interested in the energy of the crystallite relative to that of this candidate homogeneous background.  Thus we get
\beq \langle \psi | H | \psi \rangle - E_F  = (\epsilon_C - \epsilon_F) V_D + t A_D + \bar{t} A_{\bar{D}} - \epsilon_F V_S . \eeq   The first term is the expected positive volume energy of the crystal.  Recall that in the regime of interest, QMC shows that $\epsilon_F < 0$ .  Thus, the last term is positive.  For $\Delta \ll R$ the last three terms give $\Omega_{d-1} [t R^{d - 1} + \bar{t} (R + \Delta)^{d- 1} + |\epsilon_F| R^{d - 1} \Delta] $.  If we try to find a stationary point for $\Delta$ we get a solution incompatible with $\Delta \ll R$.  That is, given the approximations above, which were motivated by assuming a relatively large value for $\Delta$ in Bohr units, $\Delta$ is minimized at zero.  In reality this means that it is minimized at a microscopic value.   Thus, we have three contributions to the surface tension, two positive and one negative.  Estimating the sign of the sum is beyond our current calculational abilities.

 We now want to review an argument due to Kivelson and Spivak\cite{kivspiv}(KS), that transfer of charge from the fluid to the crystallite introduces a long range negative contribution to the surface tension, which scales like ${\rm ln} (R/R_0)$ in the two dimensional case. We write a classical Hamiltonian  for two fields, an order parameter $\phi$ and the excess charge density $\Delta n(\vec{x}) $
\beq H_{mf} =& \int d^dx[ \frac{1}{2} (\nabla \phi)^2 + U_0 (\phi (\vec{x})) + \Delta n(\vec{x}) U_1 (\phi (\vec{x}))]  \\&+ \frac{g^2}{2} \int \int d^dx d^dy \Delta n(\vec{x}) \Delta n(\vec{y}) V(\vec{x} - \vec{y}) , \eeq where $V$ is the Coulomb potential.  The charge density $\Delta n(\vec{x})$ is presumed small and we've kept only the Coulomb term because the Coulomb energy of an unscreened charge density is not extensive in the size $R$ of the crystallite.  Minimization of this energy with restpect to $\Delta n(\vec{x})$ gives an effective energy for $\phi$ 
\beq H_{eff} =& \int d^dx[ \frac{1}{2} (\nabla \phi)^2 + U_0 (\phi (\vec{x})) + \Delta n(\vec{x}) U_1 (\phi (\vec{x}))]\\&  - \frac{2c_d }{g^2} \int \int d^dx d^dy \nabla U_1 (\vec{x}) \nabla U_1 (\vec{y}) \\&\cdot \int d^d k\ e^{i\vec{k}\cdot (\vec{x} - \vec{y})} k^{2d - 3}   . \eeq  If we have a configuration of high energy density crystalline phase contained in a ball of radius $R$, surrounded by the meta-stable low energy density homogeneous phase, then the term involving $U_0$ gives the volume energy contribution, scaling like $R^d$, while the ordinary gradient term gives a positive surface tension of order $R$.  The Coulomb term gives rise to a negative non-local surface energy, which behaves at large $R$ like $R \ln (R/R_0)$ for $d = 2$ and like $g^{-2} R$ for $d = 3$.  

In two dimensions, if the shape of the crystallite is elongated, like an ellipsoid 
with axes scaling like $R$ and $R^p$ with $q = 1 + p < 2$ then the energy is
\begin{equation} E_{crystallite} = A R^q  + BR - C R {\rm ln}\ R , \end{equation} in units where the microscopic length scale $r_S$ is taken to be $1$.  A phase with crystal stripes ($p = 0$) as large as the box size $L$ is preferred over the homogeneous phase.  There are also stable, finite size crystallites for all sufficiently small $p$.  One needs more detailed calculations to tell whether a sol of finite crystallites has lower energy than the striped phase.  We refer to the extensive analysis in\cite{kivspiv} which describe the possible microemulsion phases in two dimensions.

In $3$ dimensions the long range contribution to the surface tension scales like $R$ and the question of whether charge redistribution from the fluid to the crystal leads to instability depends on microscopic physics, just as in our variational calculations.  We will comment only on one seemingly paradoxical conclusion of the Kivelson-Spivak analysis in three dimensions, which is that instability due to charge transfer seems to dominate for {\it very weak} Coulomb interactions.   We believe that this is an artifact of neglecting short range contributions to the term quadratic in $\Delta n$ in the KS energy functional.  These short range contributions are certainly there, due to the connected charge density expectation value, and they give a positive contribution to the surface tension.  In the KS functional, if the Coulomb interaction is turned off, the energy can be lowered without bound, merely by increasing $\Delta n$.   Short range quadratic terms would prevent $\Delta n$ from growing too large, and they are the dominant contributor to this inhibition, if the Coulomb interaction is too weak.

We conclude then that, in three dimensions, the existence of a sol regime must be studied with more precision.  The main reason to believe in it is the neat explanation it gives to the results of \cite{gapless}\cite{haule}.  In the next section we will propose a Landau mean field theory, based on the LDA for the density functional, which demonstrates that a homogeneous fluid with negative compressibility will have negative surface tension crystallite excitations, which lower the energy.  The ground state is a sol of these excitations. 

\section{Landau Mean Field Theory for Fluid to Colloid Transitions}\label{mean}
\par
The LDA for the density functional reads
\beq F[n] =& F_0 [n] + \frac{1}{2}\int d^dx d^dy\ \frac{(n (\vec{x}) - n_0)(n(\vec{y}) - n_0)}{|\vec{x} - \vec{y}|} \\&+ \int d^dx\ \epsilon_F(n(\vec{x})) . \eeq  Here $n_0$ is the background positive charge density and $\epsilon_F (n)$ is the energy of the lowest lying homogeneous state of the HEF, at density $n$.  The first two terms cannot have a saddle point lying on a path in $n(\vec{x})$ space between a meta-stable crystal and a stable homogeneous phase (or vice versa) because they are both convex functionals of $n(\vec{x})$ \cite{heginscol}.  A regime of negative compressibility is one over which $\epsilon_2$, the second derivative of $\epsilon_F$, is negative at the homogeneous stationary point $n = n_0 + (volume)^{-p}$.  Let us assume that the regime of negative compressibility overlaps with a regime where there exists a locally stable crystalline configuration with pronounced peaks in $n(\vec{x})$.  Since both negative compressibility {\it and} pronounced peaks, disappear at high density, it is plausible that such a regime exists.

Now set the density $n_0$ just below the point where negative compressibility sets in, and expand $n(\vec{x}) = n_0 + \Phi (\vec{x})$, around the homogeneous stationary point of the LDA functional.

 The second functional derivative of $F_0$ at $n_0$ is $-\nabla^2 + \mu_0 $, where $\mu_0$ is the chemical potential that gives density $n_0$ for noninteracting electrons. The quadratic energy functional for $\Phi$ is then\begin{equation} \int d^dk \Phi (\vec{k}) \Phi (-\vec{k}) (k^2 + \mu_0 + \epsilon_2 + k^{1 - d}). \end{equation}  
 To demonstrate an instability of the homogeneous fluid to a colloidal phase we introduce an ansatz
\begin{equation} \Phi (\vec{x}) = f_c (\vec{x},\vec{k_0}) g(\vec{x}/R) + S(\vec{x}/\Delta R ).\end{equation} $f_c$ is the density profile of the stable crystal. We take $f_c=\sum_{\vec{k_0}}\cos(\vec{k_0}\cdot\vec{x})$ in our calculations. It's important to note, that the $|\vec{k_0}|$ of the stable crystal is completely determined by $n_0$ and geometry once the lattice symmetry is given.It cannot be varied to lower the energy.  $g$ and $S$ are spherically symmetric functions. $g$ is approximately $1$ for $x < R$ and vanishes rapidly for $x > R$. $S$ vanishes outside a shell of size $\Delta R<< R$ around some $r_1 > R$ . We will also take $|\vec{k_0} R| >> 1$. In Appendix \ref{B} we show that in order for the energy of the configuration to scale like $R^d$ for large $R$ the crystal term and the shell term must both have zero charge (i.e. their $k = 0$ values must both vanish).More precisely, their total charge must go to zero rapidly enough with $R$.It is sufficient for the Fourier transforms to be very small in the region $k \sim 0$.  The energy of the crystal term by itself then scales like $R^d$ when $R$ is large, and if $k_0^2 + k_0^{1 - d} + \mu_0+ \epsilon_2 > 0$ , is positive. We show in Appendix \ref{A} that the corrections are down by two powers of R. The energy of the shell term by itself scales like $R^{d - 1}$ and it can be negative if the Fourier transform of $S$ is concentrated in the region where $k^2 + k^{1 - d} + \mu_0 + \epsilon_2$ is negative.  If the support in k space of R and the crystal term are separated from each other, then the cross terms are negligible.  So we've found a negative energy localized excitation of the fluid ground state if $\mu_0 + \epsilon_2$ is negative. 

An example of functions that do the job are $g = e^{- x^2 / 2R^2}$, $ S = \cos ( \vec{k_1} \cdot \vec{x} ) \int d^{d - 1} \hat{y} e^{- (\vec{x} - \vec{y})^2 /2 \Delta R^2} $ , with $|\vec{k_1}|\gg \Delta R^{-1} \gg |\vec{k_0}| \gg R^{-1}$, where the integral is over the direction of the vector $\vec{y}$, and $|\vec{y}|=R$.  Since we have found localized negative energy configurations, a finite density of such localized excitations will have lower energy density than the homogeneous fluid.

Of course, in the quadratic approximation, the energy is unbounded below if the compressibility is negative. We can expand the crystalline region and lower the energy. The cubic term bounds this for one sign of the deviation but not the other.  As shown long ago by Landau\cite{landau} the cubic term picks out the triangular lattice structure in two dimensions, and the face centered cubic in three.  Coincidentally, these are the same structures preferred by the low density approximation to the HEF.   Quartic terms in the expansion of the density functional will prevent the expansion of the crystal.  Recall also that the size of the crystalline region is bounded on a torus, because the periodic state guaranteed by our symmetry analysis will be a periodic superposition of states of crystal with a fluid bubble inside it.  So it does not make sense to have too large a value of $R$. 

This computation shows that if the compressibility is negative, then the homogeneous fluid is unstable to nucleation of a "sol" of crystallites . Presumably the crystallites repel each other.In order for this to be true their dominant multipole moment must be a quadrupole (dipoles can always orient themselves to attract). This will be a long range force so the colloid will stabilize at some finite density of crystallites. A true colloid would have many more or less stable configurations of similar energy. Note that given $k_0 R >> 1$, these crystallites contain many electrons and will have low energy internal phonon and rotational excitations. The effective action for the positional collective coordinate of the solution will have small coefficient so that the multibody dynamics of the crystallites can be treated semiclassically, as we have done above.Quantum mechanics will imply that the true ground state is a superposition of different configurations of the multi-crystallite positions.  

As we go really near the phase transition, $k_0 R$ goes to zero. The crystallites shrink and their energy goes to zero . They should be treated as Bose particles since they're excitations of the density operator, with no long range fields. 

It is unlikely that this mean field theory is an accurate description of the region near the transition, but it might be the basis for finding a description of the critical fluctuations, taking higher order terms in $\Phi$ into account and treating it as a quantum field.  We will reserve investigation of that question for a future paper.  

\section{Summary and Conclusions}\label{sum}
\par
We have argued, on the basis of existing QMC calculations, and other numerical evidence\cite{gapless}\cite{haule} that in two and three dimensions the HEF has a colloidal phase extending from $r_{WC}$ to a rather high density, probably close to the critical points found by\cite{gapless}\cite{haule}.
The QMC calculations show us that for $\sim 100$ electrons a homogeneous ground state is energetically favored over the WC for $r_S < r_{WC}$.  We then exhibited a large (in the volume of the system, $L^d$) number of meta-stable density configurations, consisting of large bubbles of fluid separated by thin crystalline walls, whose energy is higher than that of the homogeneous fluid by amounts of only $L^{d - 1}$.  Each of these configurations corresponds to an exponentially large Hilbert space of quantum states.  If the subsets of such states between which the Hamiltonian has matrix elements of order $1$ is $D > L^{2 d - 2}$, then random matrix theory tells us that superpositions of these states will have lower energy than the homogeneous fluid.

This very general argument used only symmetries, the existence of a neutrally stable bubble of fluid inside the crystal, and continuum elasticity theory.  The latter was justified by the macroscopic size of the neutrally stable bubble, near the point where the crossover between the energies of homogeneous crystal and fluid states occurs.
It suggests that most quantum melting transitions, the melting of the WC in particular,  will be transitions to a colloidal state.   As the control parameters are moved away from the crossover point the colloid evolves to consist of more fluid and less crystal, and passes through a conductor/insulator transition when a fluid bubble stretches across the entire volume.  Subsequent to this the system should be viewed as a sol of crystallites immersed in fluid.   The crucial question then becomes whether these crystallites have positive or negative surface tension.  In the first case, they quickly melt away to homogeneous fluid.  For negative surface tension the crystallites are stable, and the sol has lower energy than the fluid.

Using variational wave functions, we exhibited possible sources of negative surface tension for crystallites in the HEF, but were unable to prove that the surface tension was negative in general. We then reviewed a mean field argument of Kivelson and Spivak\cite{kivspiv}, which shows that transfer of charge from the fluid to the crystallite leads to a non-local negative surface tension in two dimensions, which implies that a sufficiently large and aspherical crystallite has negative surface tension, so that a sol of such crystallites has lower energy density than the homogeneous fluid.  In particular, there is always a striped phase with lower energy than the homogeneous fluid.  We note that in two dimensions, even if there are parameter regimes in which the striped phase has lower energy than a sol of crystallites, the sols are likely to be meta-stable, with rather long lifetimes, against transition to the striped phase.   If we start near the WC transition with a gel ground state, as we have argued is likely, then "adiabatic" increase of the background charge density, through the conductor/insulator transition, is likely to lead to a meta-stable sol state rather than to the more ordered stripes.

We should also note that the Hartree-Fock calculations of Overhauser\cite{overhauser}  show that a state with spin density wave order has lower energy density than the homogeneous fluid in three dimensions.  The transition occurs in a regime of $r_S$ that is of the same order as the transitions found in \cite{gapless}\cite{haule}.  We believe that, as in the case of KS stripes in two dimensions, this might also be an indicator of a-spherical crystallites with negative surface tension and energy, and thus of a sol phase.

Finally, we presented a mean field theory for second order transitions of a sol into a 
homogeneous fluid.  The key feature of the mean field theory was negative compressibility of the fluid.  Negative compressibility is a feature long known from QMC, and also appears in the calculations of \cite{gapless}\cite{haule} just below the density at which the homogeneous fluid develops a gapless mode in the charge density channel, with non-zero wave number.  Our speculative model for this transition is that the crystallites of sol, which can be thought of as "soliton" solutions of our mean field theory, shrink to zero size, zero energy Bose particles, and that the dynamics of the transition will require us to solve the mean field theory as a non-perturbative quantum field theory.

Our results leave lots of open questions unanswered.  One would like to find a lower bound on the number of quantum states sharing a roughly similar density profile, which would make our argument for the existence of a gel phase rigorous.  One would like to understand the fluctuation corrections to our mean field theory for the fluid-colloid transition.   Most importantly, it seems likely that the existence of the colloidal phase of the HEF has implications for density functional theory as it applies to real materials.  The electron density in most real materials is neither in the high density fluid regime nor in the WC regime.  In colloidal regimes the density functional is basically a probability distribution for different configurations of a classical colloid, and we imagine that this insight could lead to improved calculations of the properties of materials.

\acknowledgements TB would like to thank S.Kivelson and B.Spivak for extremely enlightening discussions about their work.  This research was supported by DOE Grant No.DOE-SC0010008.

\appendix
\section{The crystal bulk doesn't contribute to the surface energy}\label{A}\par
 Consider a crystal with an arbitrary shape. $\phi(\vec{x})=f_c(\vec{x},{\vec{k_0}})g(\vec{x}/R)=\sum_{k_0}e^{i\vec{k_0}\cdot\vec{x}}g(\vec{x}/R)$, where $g(\vec{x}/R)$ determines the overall shape of the crystal chunk. It vanishes quickly outside the crystal region. The Fourier transformation is a convolution $\phi(\vec{k})=R^d\int d^dp\sum_{k_0}\delta^d(\vec{k}-\vec{p}-\vec{k_0})\tilde{g}(R\vec{p})=\sum_{k_0}\tilde{g}(R(\vec{k}-\vec{k_0}))$, where $R^d\tilde{g}$ is the Fourier transform of $g$. The energy functional is 
\begin{equation}
E=\int\frac{d^dk}{(2\pi)^d}|\phi(\vec{k})|^2h(k)
\end{equation}
$h(k)=\frac{2\pi}{k}$ if we consider the Coulomb energy in 2D. In general, $h(k)$ is a smooth function of $k$. Denote the crystal lattice spacing as $l$, so $k_0l=2\pi$. The crystal size $R$ should satisfy $R\gg l\rightarrow k_0 L\gg 2\pi$. So $\tilde{g}(R\vec{p})$ is exponentially small for $p$ as large as $k_0$. $\tilde{g}(R\vec{p})$ should be highly concentrated at $\vec{p}=0$, so the different $k_0$ sectors decouple.
\beq
E=&R^d\sum_{k_0}\int\frac{d^dk}{(2\pi)^d}|\tilde{g}(R(\vec{k}-\vec{k_0}))|^2h(k)\\=&R^d\sum_{k_0}\int\frac{d^dp}{(2\pi)^d}|\tilde{g}(R\vec{p})|^2h(\vec{p}+\vec{k_0})
\eeq\par
Denote $\vec{q}=R\vec{p}$. 
\begin{equation}
E=R^d\sum_{k_0}\int\frac{d^dq}{(2\pi)^d}|\tilde{g}(\vec{q})|^2h(\vec{q}/R+\vec{k_0})
\end{equation}
$\tilde{g}(\vec{q})$ is concentrated at $\vec{q}=0$. We can do an expansion in terms of $\vec{q}/R$. 
\begin{equation}
E=R^d\sum_{k_0}\int\frac{d^dq}{(2\pi)^d}|\tilde{g}(\vec{q})|^2\sum_{n=0}^{\infty} \frac{1}{n!}(\vec{q}\cdot\nabla_{\vec{k_0}}/R)^nh(k_0)
\end{equation}
Since $|\tilde{g}(\vec{q})|^2$ is an even function of $\vec{q}$, all odd orders should be 0. 

\section{The crystal bulk and the shell must both have zero net charge}\label{B}\par
If not so, we can separate the charge distribution into two parts: $\rho(\vec{x})=\rho_b(\vec{x})+\rho_p(\vec{x})$, where $\rho_b(\vec{x})$ is the background with nonzero net charge in both the crystal region and the shell region, while $\rho_p(\vec{x})$ is the periodic part with zero net charge in these two regions. As the configuration expands, its radial size goes from $r$ to $\lambda r$,  $\rho_b(\vec{x})$ becomes $\rho'_b(\vec{x})=\rho_b(\vec{x}/\lambda)$. Its Fourier transform is
\beq
\rho'_b(\vec{k})&=\int d^dx \rho'_b(\vec{x})e^{i\vec{k}\cdot \vec{x}}
=\lambda^d\int d^d\frac{x}{\lambda}\rho_b(\vec{x}/\lambda)e^{i\lambda\vec{k}\cdot \vec{x}/\lambda}\\&= \lambda^d\rho_b(\lambda\vec{k}) 
\eeq
The Coulomb energy contributed by $\rho'_b(\vec{x})$ is
\beq
E'_C &\propto \int d^dk |\rho'_b(\vec{k})|^2 k^{1-d}
\\&=\lambda^{2d-1}\int d^d(rk)  |\rho_b(\lambda\vec{k})|^2(\lambda k)^{1-d}\propto \lambda^{2d-1}E_C
\eeq
So the Coulomb energy of $\rho_b$ scales like $\lambda^{2d-1}\neq \lambda^d$ when $d>1$.  The total Coulomb energy of the whole configuration will not be extensive, either. So the crystal and the shell must both have zero net charge.

\section{The $\sigma/r_d - \epsilon$ plot in 2D}
\label{C}
\begin{figure}[H]
\centering
\includegraphics[scale=0.55]{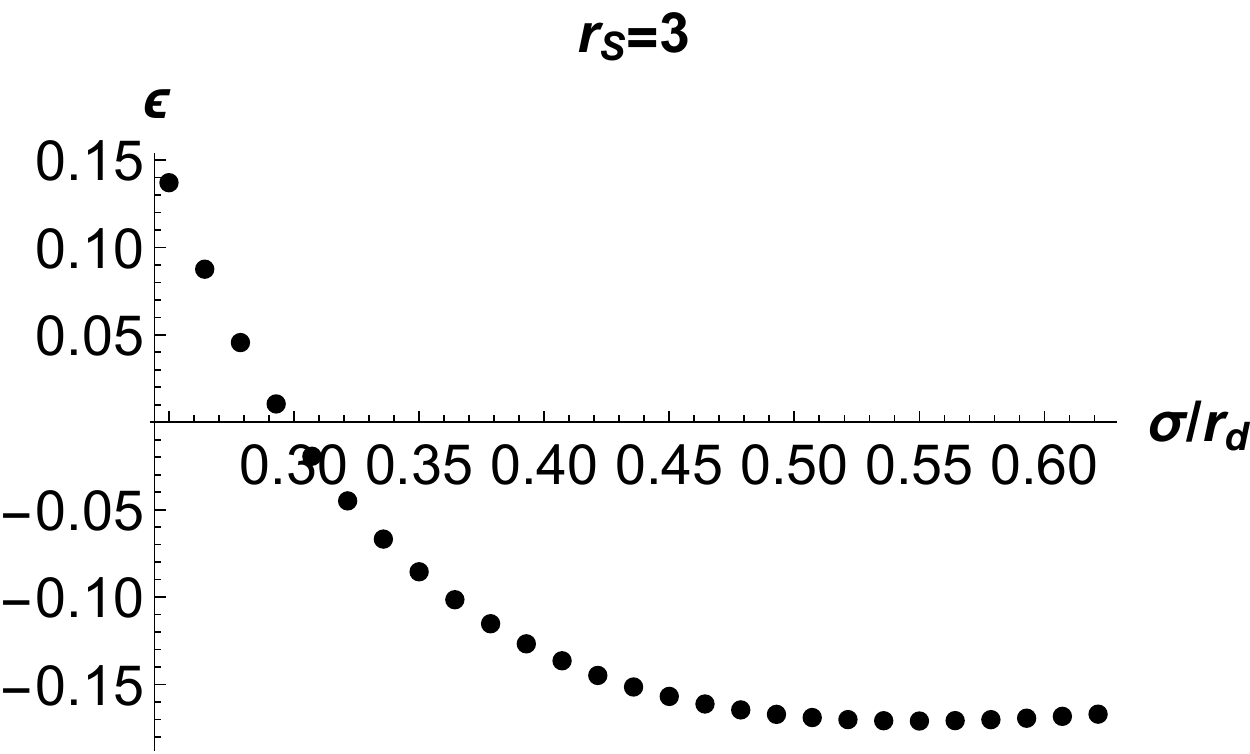}
\end{figure}
\begin{figure}[H]
\centering
\includegraphics[scale=0.55]{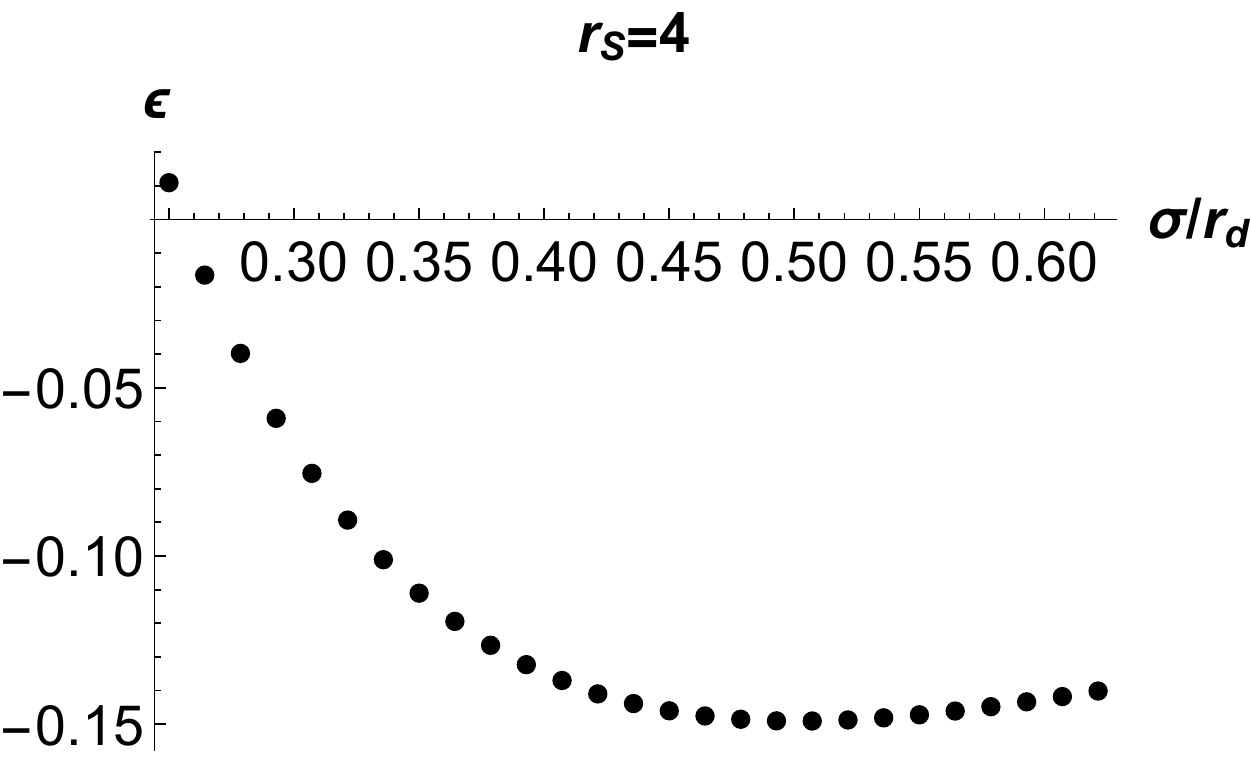}
\end{figure}
\begin{figure}[H]
\centering
\includegraphics[scale=0.55]{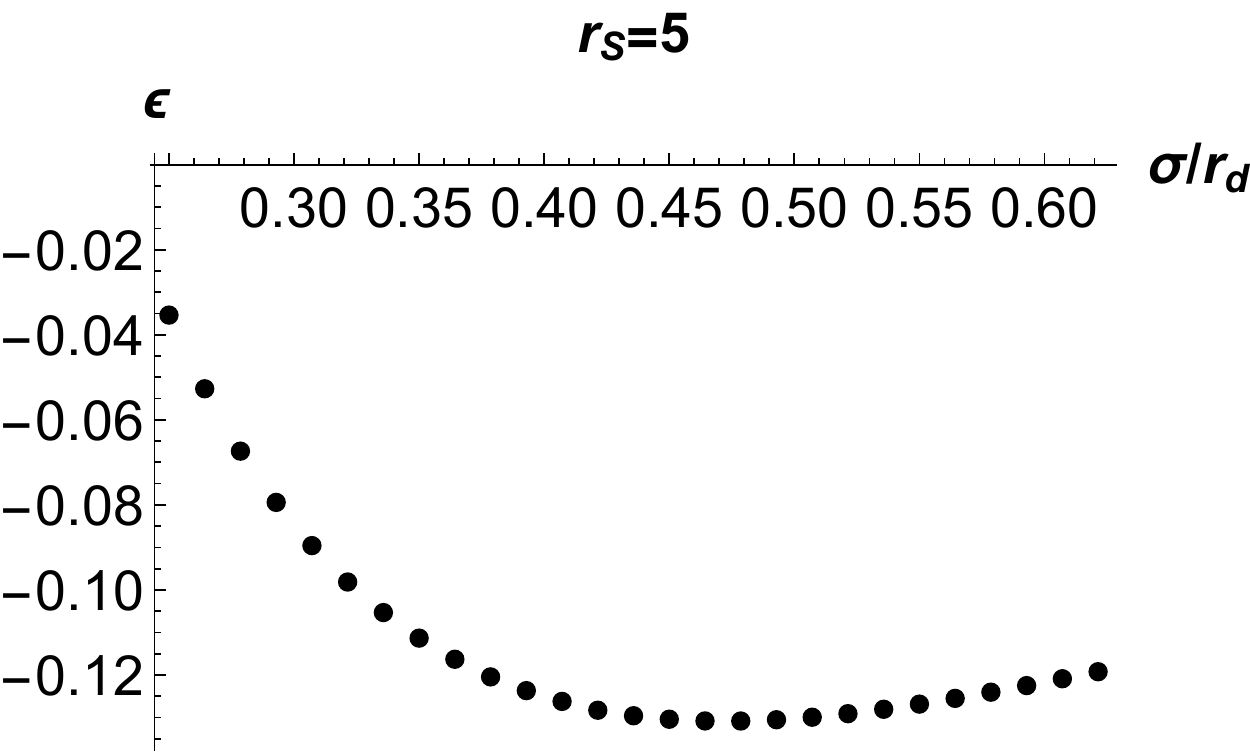}
\end{figure}
\begin{figure}[H]
\centering
\includegraphics[scale=0.55]{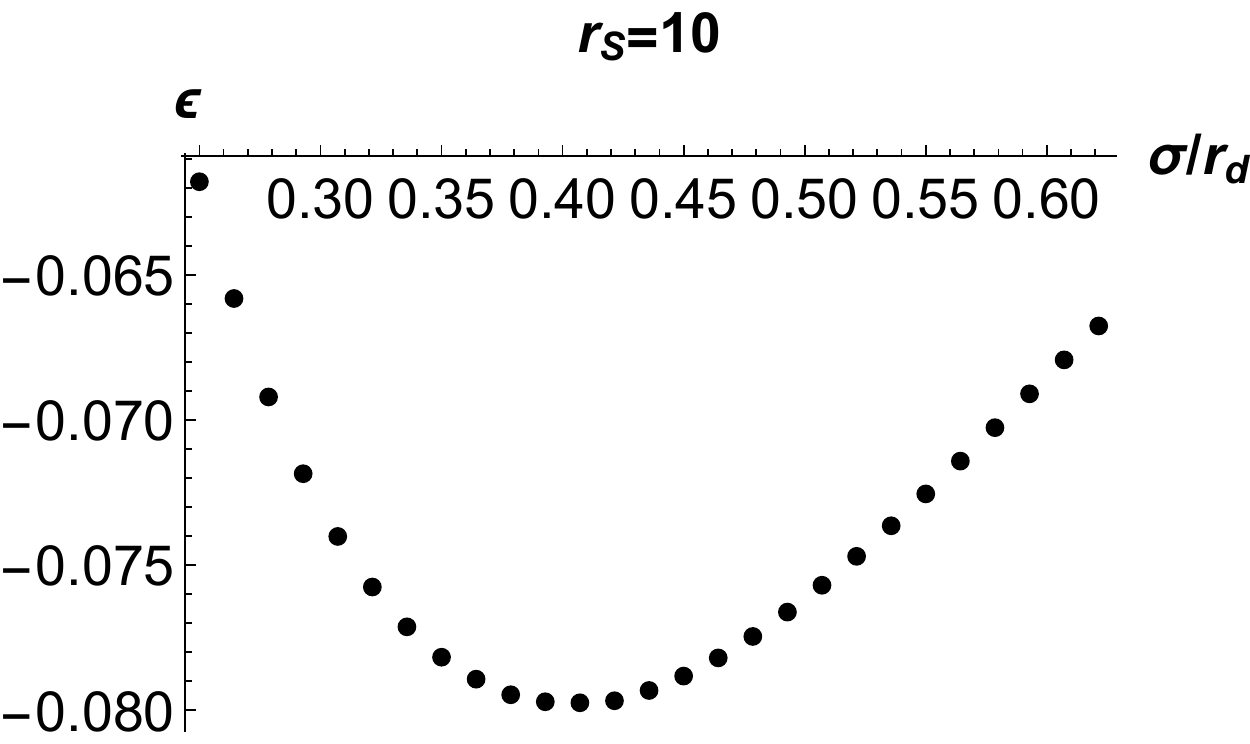}
\caption{$\sigma/r_d-\epsilon$ plot in 2D, where $r_d=\sqrt{\frac{\pi}{\sqrt{3}}}r_S$ is the distance between two lattice points on the two dimensional triangular lattice. A saddle point appears within the $\sigma/r_d\leq 0.5$ region when $r_S\geq 4$.}
\label{fig1}
\end{figure}
\leavevmode
\section{The $\sigma/r_d - \epsilon$ plot in 3D}
\label{D}
\begin{figure}[H]
\centering
\includegraphics[scale=0.55]{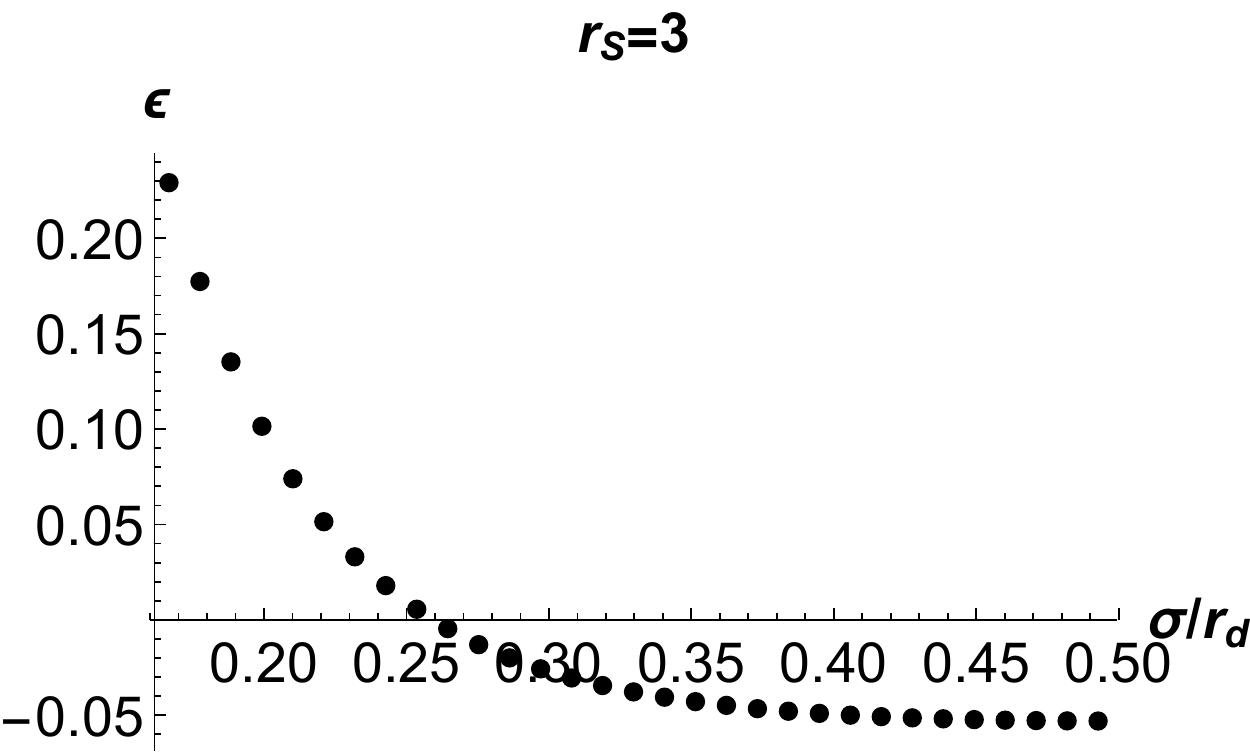}
\end{figure}
\begin{figure}[H]
\centering
\includegraphics[scale=0.55]{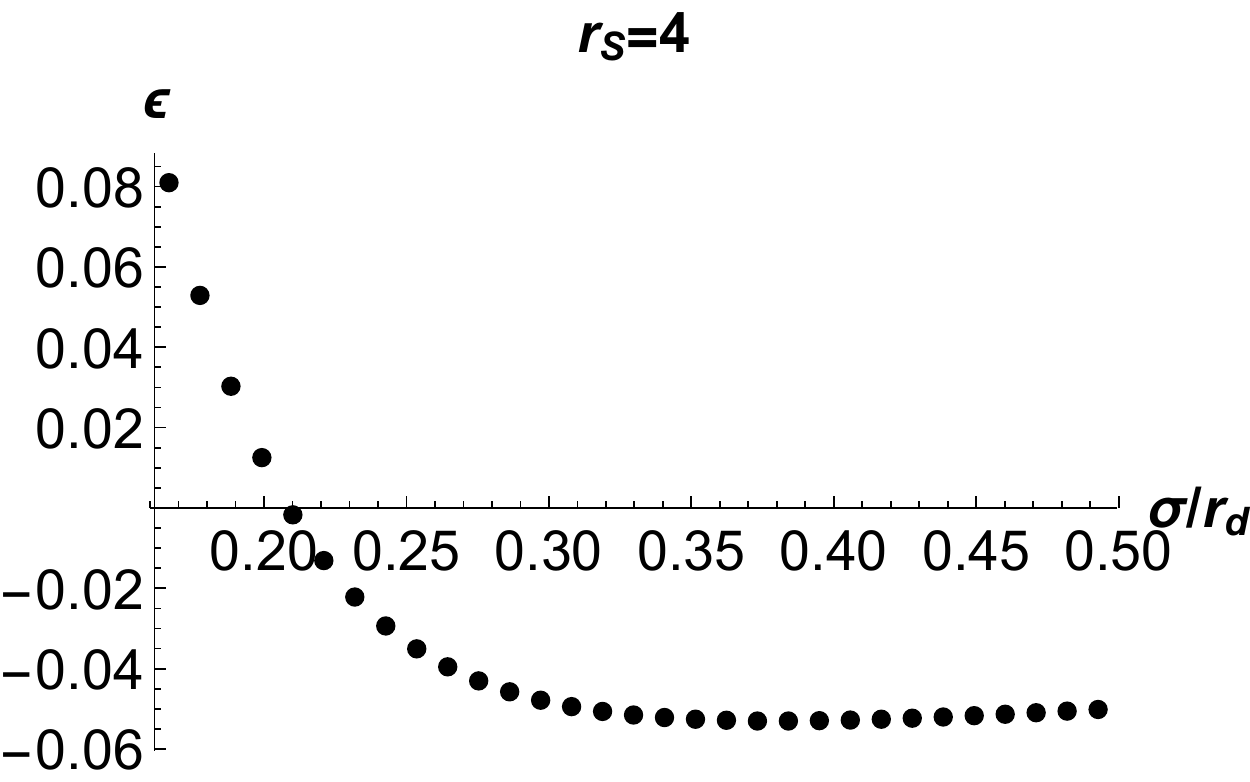}
\end{figure}
\begin{figure}[H]
\centering
\includegraphics[scale=0.55]{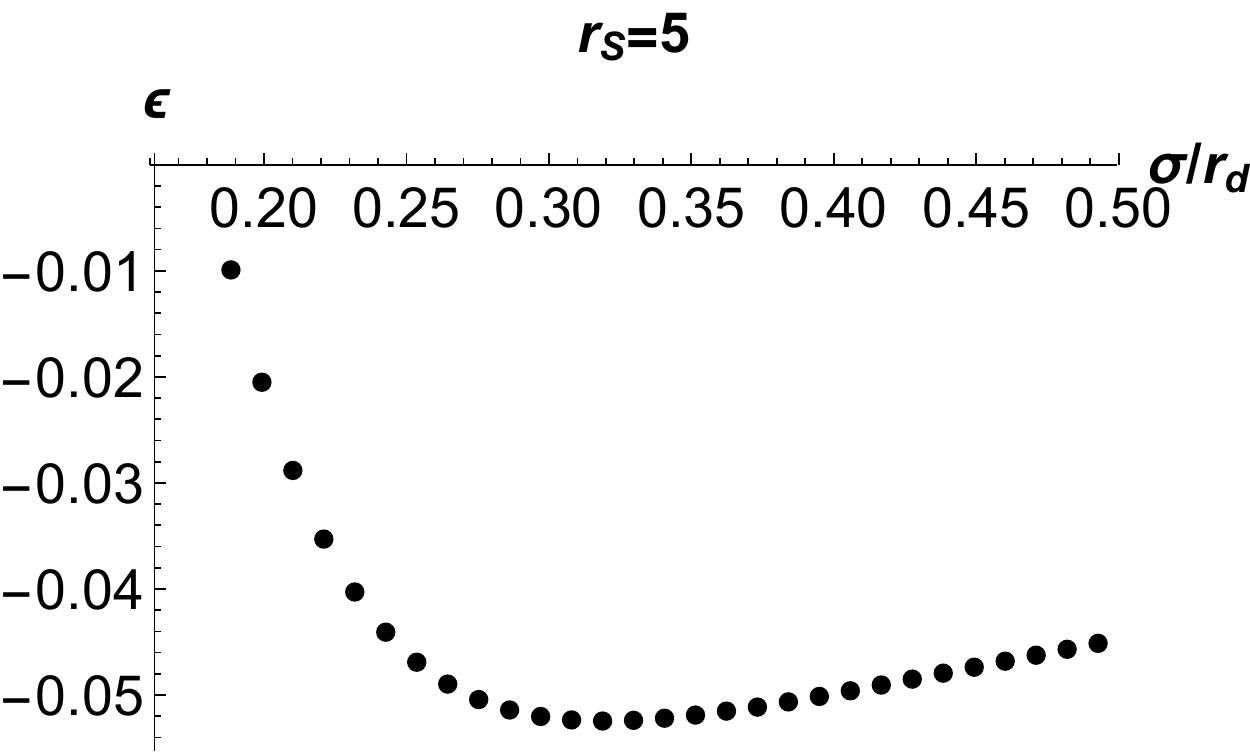}
\end{figure}
\begin{figure}[H]
\centering
\includegraphics[scale=0.55]{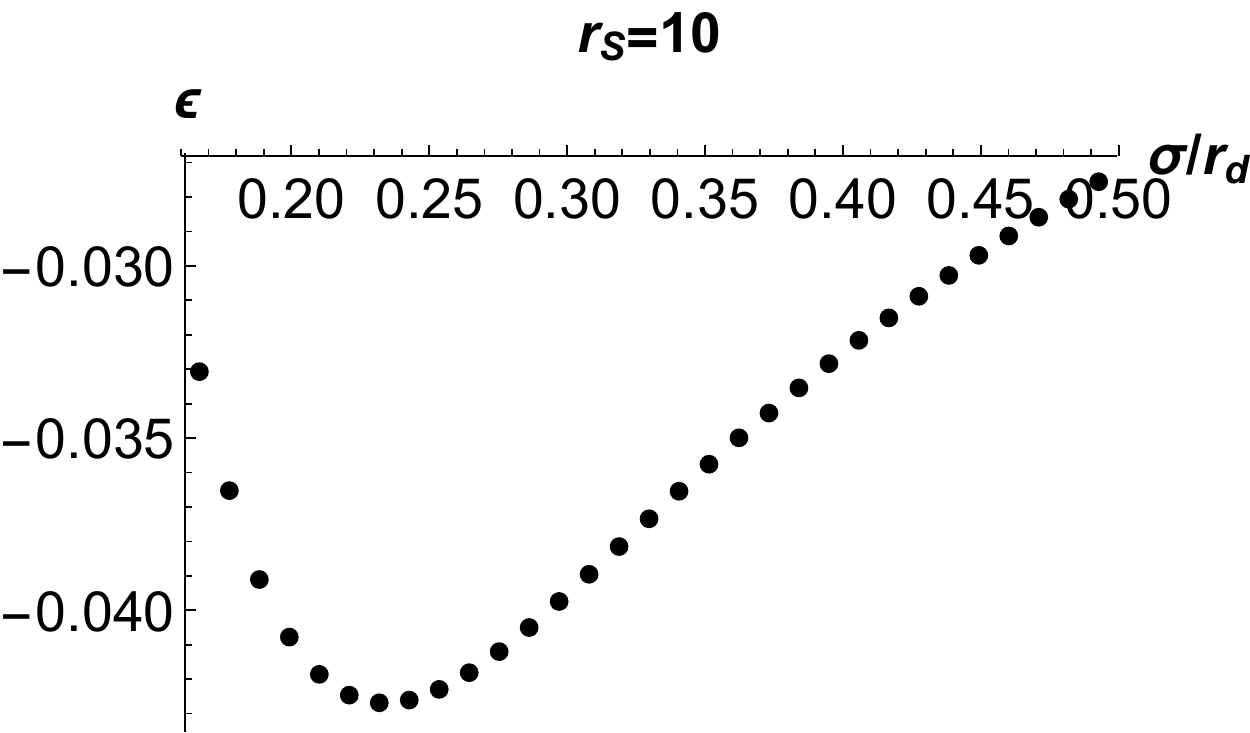}
\caption{$\sigma/r_d-\epsilon$ plot in 3D, where $r_d=\sqrt{3}(\frac{\pi}{3})^{1/3}r_S$ is the distance between two lattice points on the bcc lattice. A saddle point appears within the $\sigma/r_d\leq 0.5$ region when $r_S\geq 4$.}
\label{fig3}
\end{figure}

\end{document}